\documentclass{article}

\newcommand{\inst}[1]{\textsuperscript{#1}}

\usepackage[pages=some]{background}
\backgroundsetup{
scale=1,
color=black,
opacity=0.4,
angle=0,
contents={%
  \includegraphics[width=\paperwidth,height=\paperheight]{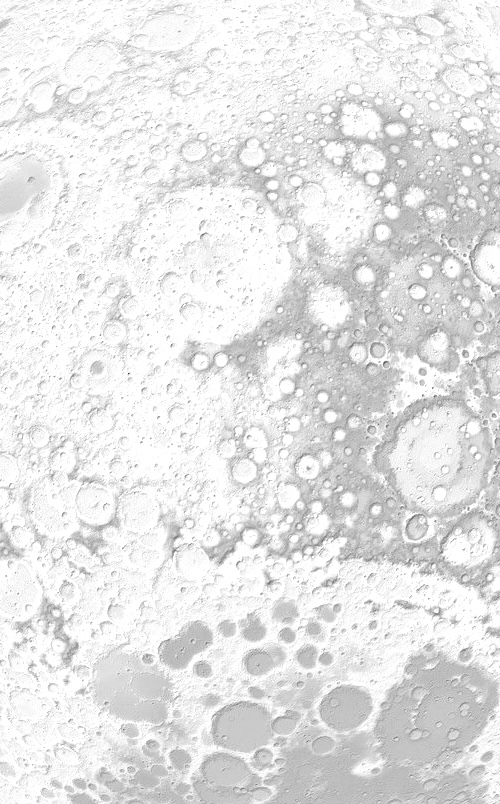}
  }%
}

\usepackage{wrapfig}
\usepackage[absolute,overlay]{textpos}
\usepackage{siunitx}
\usepackage{xcolor}
\usepackage{upgreek}
\usepackage{comment}
\usepackage{sectsty}
\usepackage{multicol}
\definecolor{ESO_blue}{RGB}{1, 132, 184}
\definecolor{gainsboro}{RGB}{220, 220, 220}
\definecolor{platinum}{RGB}{229, 228, 226}
\definecolor{white_smoke}{RGB}{245, 245, 245}
\sectionfont{\color{ESO_blue}\fontsize{13pt}{16pt}\selectfont}
\subsectionfont{\color{ESO_blue}\fontsize{12pt}{16pt}\selectfont}

\usepackage{titlesec}
\titlespacing*{\section}
{0pt}{0.5ex plus 0.2ex minus 0.1ex}{0.3ex plus 0.1ex}
\titlespacing*{\subsection}
{0pt}{0.8ex plus .5ex minus .2ex}{0.6ex plus .2ex}

\usepackage{natbib}
\usepackage[labelfont=bf, textfont=normalfont]{caption}

\makeatletter
\renewcommand{\@seccntformat}[1]{}
\makeatletter

\usepackage{graphicx}
\begin{document}
\fontsize{11pt}{13pt}\selectfont

%------------------------------------------------------------------------------
% TITLE PAGE
\begin{titlepage}
    \BgThispage
    \centering
    \vspace*{4cm}

    % ---- Title ----
    {\Huge\bfseries\color{ESO_blue} Hunting exomoons with a kilometric baseline interferometer\par}
    \vspace{0.7cm}
    {\Large\bfseries\color{ESO_blue} ESO Expanding Horizons White Paper\par}
    \vspace{2cm}
    % ---- Authors ----
    {\Large
    Thomas\,O.\,Winterhalder\inst{1,2},
    Antoine\,M\'erand\inst{1},
    Sylvestre\,Lacour\inst{1,3},
    Jens\,Kammerer\inst{1},
    Guillaume\,Bourdarot\inst{4},
    Frank\,Eisenhauer\inst{4}
    }
    
    \vspace{2cm}
    % ---- Affiliations ----
    {\normalsize
    \textsuperscript{1} European Southern Observatory, Karl-Schwarzschildstrasse 2, D-85748 Garching bei München, Germany \\
    \textsuperscript{2} Leiden Observatory, Leiden University, P.O. Box 9513, 2300 RA Leiden, The Netherlands \\
    \textsuperscript{3} LIRA, Observatoire de Paris, Universit\'e PSL, CNRS, Sorbonne Universit\'e, Universit\'e de Paris, 5 place Jules Janssen, 92195 Meudon, France \par
    \textsuperscript{4} Max-Planck-Institut für extraterrestrische Physik, Giessenbachstra\ss e~1, 85748 Garching, Germany\label{mpe} \par
    }

    \vfill
    {\large \today\par}

    \begin{textblock*}{7cm}(14.8cm,29cm)
        {\tiny\color{gray} Credit: NASA/Goddard Space Flight Center/DLR/ASU}
        % Full source: https://science.nasa.gov/resource/high-resolution-topographic-map-of-the-moon/
    \end{textblock*}
    
\end{titlepage}

%------------------------------------------------------------------------------
\pagecolor{white}
\section{Abstract}
\textbf{Despite numerous search campaigns based on a diverse set of observational techniques, exomoons -- prospective satellites of extrasolar planets -- remain an elusive and hard-to-pin-down class of objects. Yet, the case for intensifying this search is compelling: as in the Solar System, moons can act as proxies for studying planet formation and evolution, provide direct clues as to the migration history of the planetary hosts and, in favourable cases, offer potentially habitable environments. Here, we present an investigation into how the search for exomoons would benefit from a new interferometric facility operating in the optical wavelength domain and leveraging baselines substantially longer than the ones the VLTI is currently equipped with.
We find that an interferometer providing an astrometric precision of 1\,$\upmu$as would be able to robustly detect Earth-mass and sub-Earth-mass exomoons on dynamically stable orbits around Jupiter-like planets at distances between 50 and 200\,pc.
}

\section{Why pursue exomoons?}
Our current understanding of planet formation and evolution is largely driven by theoretical models that manage to capture some of the population-level trends and features emerging from an ever-expanding sample of observations (Mordasini et al. 2009). Constraining the formation history of a given planet, on the other hand, remains challenging.
In the future, exomoons can serve as an additional proxy as to the dynamical history of the host planet. For instance, the sheer presence of the Galilean moons together with the fact that each of them revolves around Jupiter on an almost circular, coplanar orbit suggests that they formed in situ within the circumplanetary disc and that the planet cannot have undergone any strong gravitational encounters or eccentric migration (e.g. Canup \& Ward 2002). In this way, the presence and orbital geometry of extrasolar moons can inform us about the evolution of their planetary hosts.
On top of this, a small subset of exomoons can conceivably provide a habitable environment for life to thrive (e.g. Heller et al. 2013). Even in our own Solar System, the icy moons of Jupiter and Saturn are believed to harbour liquid water in vast subsurface oceans, despite the fact that they are located far beyond what is traditionally considered the habitable zone (Carr et al. 1998; Kivelson et al. 2000). Apart from tidally heated moons around far-out planets like Europa and Ganymede, moons around gas giants within the aforementioned habitable zone of their respective system are another potentially clement class of objects.

While we are still awaiting the first unambiguous detection of an extrasolar moon, the above examples serve as an indication of how their eventual discovery and characterisation will impact the field of exoplanet science and the search for life in the future.

\section{Astrometric detection}
There exists a multitude of different techniques proposed to detect moons around exoplanets. Among them are methods based on planet and, in some cases, moon transit observations (Szab{\'o} et al. 2006; Limbach et al. 2021), radial velocity time series data (Ruffio et al. 2023; Vanderburg \& Rodriguez 2021), microlensing events (Liebig \& Wambsganss 2010) as well as direct moon imaging (Lazzoni et al. 2020).
Here, we will concentrate on an altogether different method.
%In the same way that a planet in orbit around a star induces reflex motion in its host, a given moon will cause orbital motion of its planetary host around the planet-moon centre-of-mass.
The presence of an orbiting moon gravitationally perturbing a given planet in orbit around a star can manifest itself as a low-amplitude epicyclic motion of the planetary host that -- given a sufficiently precise instrumental setup -- can be resolved and measured.
Repeated measurements of the position of the planet relative to the star offer a means of detecting these deviations from a perfectly Keplerian two-body orbit (Winterhalder et al. 2025).

\begin{figure}[t]
    \centering
    \includegraphics[width=\textwidth]{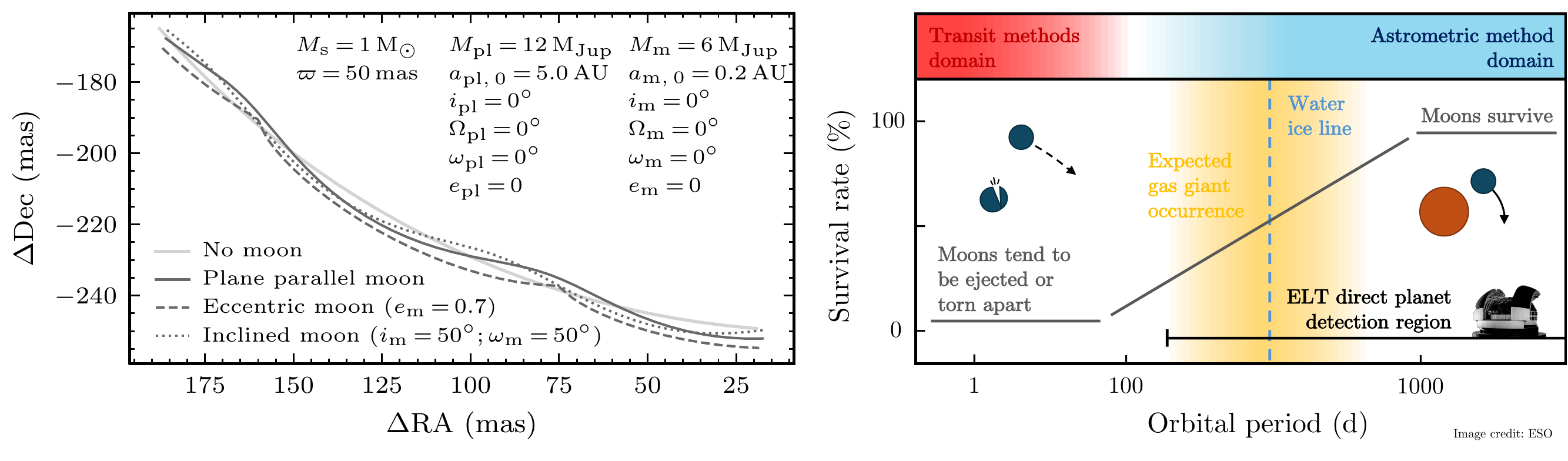}
    \caption{\textit{Left:} Schematic of the orbital wobble exhibited by a planet as a result of being orbited by a moon. The system parameters in the top right indicate the baseline `plane parallel moon' case. The adjustments made for the alternative configurations are indicated in brackets in the bottom left. This is a reproduction of Fig.~1 in Winterhalder et al. (2025). \textit{Right:} Schematic loosely sketching the parameter space accessible using the astrometric technique as compared to the transit methods of exomoon detection in the context of the moon survival rate as a function of planetary orbital period. Additionally, the expected peak of the gas giant occurrence around the water ice line as well as the predicted ELT planet detection range are indicated.}
    \label{figure_schematic_combined}
\end{figure}

Any planetary targets selected for an astrometric exomoon hunt must be amenable to direct imaging techniques. For instrumental reasons, these are mostly young, self-luminous planets on wide orbits.
% I THINK THIS IS NOT TRUE: MOONS AROUND FAR OUT PLANETS WOULD STILL PRODUCE A MEASURABLE REFLEX MOTION IN THE PLANET
%The nature of the astrometric technique imposes another requirement, however: the need for resolving the perturbations of a potential moon in the trajectory of the planet implies that orbital periods should be sufficiently small such that deviations can be detected within a reasonable amount of time.
Although this excludes planets with periods of less than a few hundred days, the parameter space at which the astrometric method is applicable coincides with the expected peak in the giant planet occurrence rate around the water ice line, providing a large reservoir of planets that can be probed for the presence of an orbiting moon (Fernandes et al. 2019; Fulton et al. 2021). Additionally, the domain in which the method can be utilised overlaps with the steep increase in moon survival rate as suggested by numerical simulations (e.g. Dobos et al. 2021). These imply that at shorter periods and hence smaller separations to the star the planetary Hill radius shrinks so severely that any gravitationally bound moon is likely to either be ejected or torn apart (e.g. Barnes \& O'Brien 2002).
The astrometric method thus allows us to probe a very specific parameter space, which is favourable in terms of both the number of probable planets and the likelihood of their moons surviving over long timescales, as well as inaccessible to methods that hinge on transit observations. A visual summary of these arguments is presented in the right panel of Fig.~\ref{figure_schematic_combined}.

\section{Leveraging a kilometre-baseline interferometer}
While, in principle, any instrument capable of directly detecting a planet can be used to follow through on this strategy, the pivotal metric that governs the sensitivity of the method is the precision with which the astrometric measurements are made.
Classical direct imaging serves as an excellent means for detecting and characterising young, far-out giant planets (Bowler et al. 2016). Yet, the method is inherently limited in the astrometric precision it can provide. Measurements obtained using current state-of-the-art instruments like the VLT/SPHERE or GPI are subject to uncertainties of the order of $\SI{1}{mas}$ (e.g. Wang et al. 2016; Maire et al. 2021).
By contrast, combining telescopes separated by long baselines using optical interferometric methods can facilitate astrometric measurements that are several orders of magnitude more precise (Shao \& Colavita 1992). For instance, with Unit Telescope baseline lengths between $\SI{47}{}$ and $\SI{130}{m}$, VLTI/GRAVITY routinely obtains planet position measurements as precise as $\SI{50}{\micro as}$ (Lacour et al. 2019).
This unique capability renders optical interferometric facilities the optimal choice for monitoring the orbital movement of a given planet with the aim of detecting a perturbing moon.
In the future, a facility that leverages baselines spanning several kilometres would dramatically improve our astrometric precision, reducing measurement uncertainties to $\SI{1}{\micro as}$ (Bourdarot \& Eisenhauer 2024).
Not even the upcoming Extremely Large Telescope (ELT) can match these capabilities (Trippe et al. 2010). That being said, the ELT will aid the astrometric exomoon search by directly imaging a multitude of potentially moon-hosting gas giant exoplanets on wide orbits. In this sense, a kilometric baseline interferometer can be viewed as the ideal successor of the ELT in that it would be capable of taking the next step: detecting or ruling out moons around planets discovered and directly imaged by the ELT.

\section{Detection thresholds and sensitivity curves}
\begin{figure}[t]
    \centering
    \includegraphics[width=0.99\textwidth]{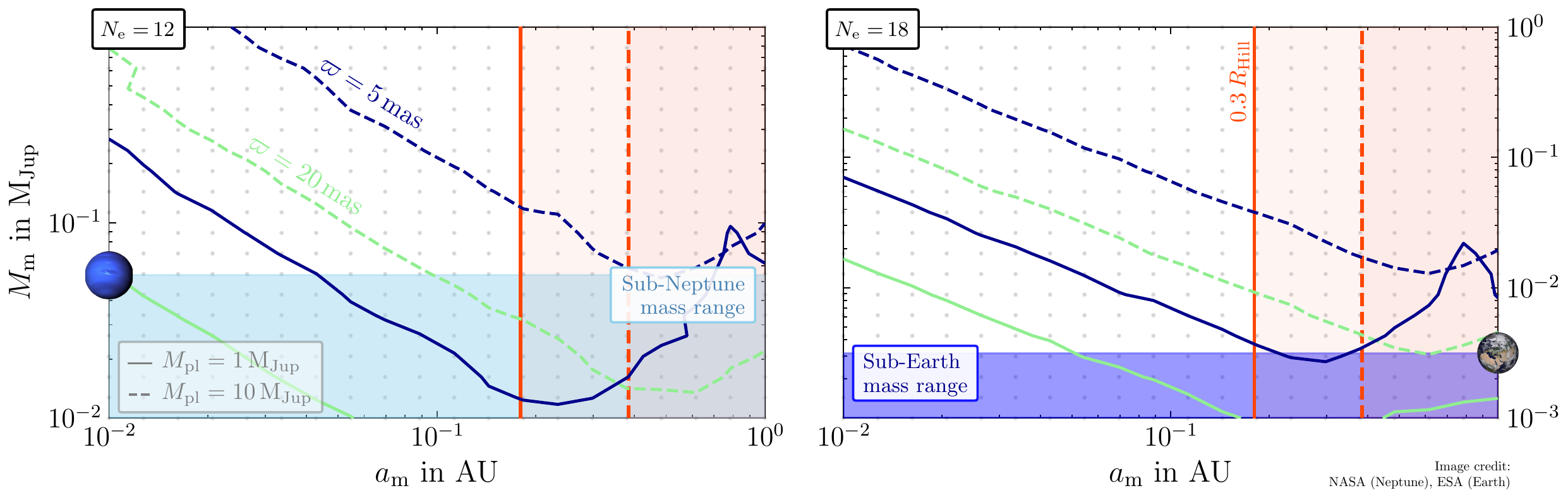}
    \caption{5$\sigma$ moon detection sensitivity curves around an exemplary exoplanet ($M_\mathrm{s}=\SI{1.5}{M_\odot}$; $a_\mathrm{pl}=\SI{10}{AU}$; $i_\mathrm{pl}=\SI{0}{^\circ}$, i.e. face-on; $e=\SI{0}{}$) using $N_\mathrm{e}=12$ and 18 astrometric epochs. The two line colours correspond to the different parallax settings of $\varpi=5$ and $\SI{20}{mas}$ as indicated in the left panel, the solid and dashed lines relate to the host planet mass settings of $\SI{1}{}$ and $\SI{10}{M_{Jup}}$, respectively. The dynamical instability domain beyond $0.3$ times the respective Hill radius, $R_\mathrm{Hill}$, is shaded red for the different planetary masses. The mass ranges below one Neptune and one Earth-mass are shaded in light and dark blue.}
    \label{figure_sensitivity_curves}
\end{figure}

The case for pursuing astrometric binary planet and exomoon detections using VLTI/GRAVITY was laid out in Winterhalder et al. (2025). Here, we focus on a more detailed investigation of the possibilities afforded by a future interferometric facility that provides an astrometric precision of $\SI{1}{\micro as}$.
By integrating the orbital trajectories of different exoplanets perturbed by an unseen moon, we can simulate the dynamical behaviour of a star-planet-moon system given varying moon masses and orbital parameters. Generating mock astrometric epochs from the fiducial orbits, we can construct realistic time series data sets as obtained by different instruments. Fitting these with a moon model and a no-moon model, we can then convert the mock epochs into an achievable detection significance. An iterative application of this procedure over a grid of different parameter settings enables us to map out our moon detection sensitivity.
Figure~\ref{figure_sensitivity_curves} shows the 5$\sigma$ moon detection sensitivity curves computed in this way for an exemplary host planet at different parallax values and for a varying amount of astrometric epochs. They show that -- up to comparatively large distances of $\SI{200}{pc}$ and on the basis of only 12 epochs -- a future kilometric baseline facility could confidently detect exomoons with masses down to a few percent of a Jupiter-mass orbiting their Jupiter-like host planet within the stable regime inside 0.3\,$R_\mathrm{Hill}$. For closer host planets at distances up to $\SI{50}{pc}$, these capabilities extend to below one percent of a Jupiter-mass.
Increasing the number of epochs to 18 (see right panel of Fig.~\ref{figure_sensitivity_curves}), moon masses down to approximately one Earth-mass can be robustly detected around Jupiter-like hosts out to distances of $\SI{200}{pc}$. For closer systems, sub-Earth-masses can be accessed around the same class of host planets.

\section{Conclusions}
In this paper, we have laid out the case for pursuing the astrometric detection of exomoons by harnessing the unique capabilities of a kilometre-baseline optical interferometer.
The method itself aims to detect the orbital wobble motion that a given planet will exhibit when hosting a moon of sufficient mass.
As opposed to transit-based methods, the astrometric technique is applicable to a parameter space where moon survival over long timescales is particularly likely: at large planet-to-star separations. 
While current instruments like VLTI/GRAVITY are sensitive to more extreme system configurations (Kral et al. 2025), a future facility providing an astrometric precision of $\SI{1}{\micro as}$ is capable of confidently detecting Earth-mass and even sub-Earth-mass moons orbiting within the dynamically stable planet-moon separation regime below $0.3$\,$R_\mathrm{Hill}$ at distances between \SI{50}{} and \SI{200}{pc}. Thus, a kilometric baseline interferometric facility would for the first time enable us to hunt for habitable exomoons around gas giant exoplanets orbiting their host stars in the habitable zone.

\vfill
%{\bf References:}
\section{References}
\begin{multicols}{2}
Mordasini et al. (2009), A\&A, 501, 1139 \\
Canup \& Ward (2002), AJ, 124, 3404 \\
Heller et al. (2013), Astrobiology, 13, 18 \\
Carr et al. (1998), Nature, 391, 363 \\
Kivelson et al. (2000), Science, 289, 1340 \\
Szab{\'o} et al. (2006), A\&A, 450, 395 \\
Limbach et al. (2021), ApJ, 918, L25 \\
Ruffio et al. (2023), AJ, 165, 113 \\
Vanderburg\& Rodriguez (2021), ApJ, 922, L2 \\
Liebig \& Wambsganss (2010), A\&A, 520, A68 \\
Lazzoni et al. (2020), A\&A, 641, A131 \\
Winterhalder et al. (2025), arXiv:2509.15304 \\
Fernandes et al. (2019), ApJ, 874, 81 \\
Fulton et al. (2021), ApJS, 255, 14 \\
Dobos et al. (2021), PASP, 133, 094401 \\
Barnes \& O'Brien (2002), ApJ, 575, 1087 \\
Bowler et al. (2016), PASP, 128, 102001 \\
Wang et al. (2016), AJ, 152, 97 \\
Maire et al. (2021), JATIS, 7, 035004 \\
Shao \& Colavita (1992), A\&A, 262, 353 \\
Lacour et al. (2019), A\&A, 623, L11 \\
Bourdarot \& Eisenhauer (2024), arXiv:2410.22063 \\
Trippe et al. (2010), MNRAS, 402, 1126 \\
Kral et al. (2025), arXiv:2511.20091
\end{multicols}

\end{document}